\documentclass[
 superscriptaddress,
 longbibliography,
 nostarredaffiliations,
 twocolumn,
 amsmath,amssymb,
]{revtex4-2}
\usepackage[utf8]{inputenc} 
\usepackage[T1]{fontenc}
\usepackage{booktabs} 
\usepackage{lmodern} 
\usepackage{graphicx}
\usepackage{dcolumn}
\usepackage{bm}
\usepackage[colorlinks=true, breaklinks=true, linkcolor=blue, citecolor=blue, urlcolor=blue]{hyperref}
\usepackage{color}
\usepackage{xcolor}
\usepackage{siunitx}
\usepackage{times}
\usepackage{braket}
\usepackage{tabularx}
\usepackage{mathtools}
\usepackage{rotating}
\usepackage{tikz}
\usepackage{newfloat}

\DeclareFloatingEnvironment[
fileext=lof, 
listname={List of SM Figures}, 
name={Supplementary Figure}, 
placement=htp 
]{SM_fig}

\makeatletter
\def\@hangfrom@section#1#2#3{\@hangfrom{#1#2}#3}
\def\@hangfroms@section#1#2{#1#2}
\makeatother

\allowdisplaybreaks

\newcommand{\dspin}{\ket{\downarrow}}

\newcommand{\nm}{\si{\nano\meter}}

\newcommand{\LSamp}{\Delta_{\text{LS}}}
\newcommand{\LSmod}{\omega_{\text{LS}}}

\newcommand{\rabiEff}{\widetilde{\Omega}}

\begin{document}

\title{Measuring spectral functions of doped magnets with Rydberg tweezer arrays}

\author{Romain~Martin$^{\ddagger}$}
\altaffiliation{These authors contributed equally to this work.}
\affiliation{
Universit\'{e} Paris-Saclay, Institut d'Optique Graduate School, CNRS, Laboratoire Charles Fabry, 91127 Palaiseau Cedex, France
}
\email{romain.martin1@institutoptique.fr}

\author{Mu~Qiao}
\altaffiliation{These authors contributed equally to this work.}
\affiliation{
Universit\'{e} Paris-Saclay, Institut d'Optique Graduate School, CNRS, Laboratoire Charles Fabry, 91127 Palaiseau Cedex, France
}

\author{Ivan~Morera}
\altaffiliation{These authors contributed equally to this work.}
\affiliation{Institute for Theoretical Physics, ETH Zurich, 8093 Zurich, Switzerland.}

\author{Lukas~Homeier}
\altaffiliation{These authors contributed equally to this work.}
\affiliation{JILA and Department of Physics, University of Colorado, Boulder, CO, 80309, USA}
\affiliation{Center for Theory of Quantum Matter, University of Colorado, Boulder, CO, 80309, USA}

\author{Bastien~G\'ely}
\affiliation{
Universit\'{e} Paris-Saclay, Institut d'Optique Graduate School, CNRS, Laboratoire Charles Fabry, 91127 Palaiseau Cedex, France
}

\author{Lukas~Klein}
\affiliation{
Universit\'{e} Paris-Saclay, Institut d'Optique Graduate School, CNRS, Laboratoire Charles Fabry, 91127 Palaiseau Cedex, France
}

\author{Yuki~Torii~Chew}
\affiliation{
Universit\'{e} Paris-Saclay, Institut d'Optique Graduate School, CNRS, Laboratoire Charles Fabry, 91127 Palaiseau Cedex, France
}

\author{Daniel~Barredo}
\affiliation{
Universit\'{e} Paris-Saclay, Institut d'Optique Graduate School, CNRS, Laboratoire Charles Fabry, 91127 Palaiseau Cedex, France
}
\affiliation{
Nanomaterials and Nanotechnology Research Center (CINN-CSIC), Universidad de Oviedo (UO), Principado de Asturias, 33940 El Entrego, Spain
}

\author{Thierry~Lahaye}
\affiliation{
Universit\'{e} Paris-Saclay, Institut d'Optique Graduate School, CNRS, Laboratoire Charles Fabry, 91127 Palaiseau Cedex, France
}

\author{Eugene~Demler}
\affiliation{Institute for Theoretical Physics, ETH Zurich, 8093 Zurich, Switzerland.}

\author{Antoine~Browaeys}
\email{antoine.browaeys@institutoptique.fr}
\affiliation{
Universit\'{e} Paris-Saclay, Institut d'Optique Graduate School, CNRS, Laboratoire Charles Fabry, 91127 Palaiseau Cedex, France
}

\date{\today}
\begin{abstract}
Spectroscopic measurements of single-particle spectral functions provide crucial insight into strongly correlated quantum matter by resolving the energy and spatial structure of elementary excitations. Here we introduce a spectroscopic protocol for single-charge injection with simultaneous spatial and energy resolution in a Rydberg tweezer array, effectively emulating scanning tunneling microscopy. By combining this protocol with single-atom–resolved imaging, we go beyond conventional spectroscopy by not only measuring the single-particle spectral function but also directly imaging the microscopic structure of the excitations underlying spectral resonances in frustrated $tJ$ Hamiltonians. We reveal resonances associated with the formation of bound magnetic polarons -- composite quasiparticles consisting of a mobile hole bound to a magnon -- and directly extract their binding energy, spatial extent, and spin character. Finally, by exploiting the spatial tunability of our platform, we measure the local density of states across different lattice geometries. Our work establishes Rydberg tweezer arrays as a powerful platform for spectroscopic studies of strongly correlated models, offering microscopic control and direct real-space access to emergent quasiparticles in engineered quantum matter.
\end{abstract}

\maketitle

\section{Introduction}

Understanding the nature of elementary excitations in correlated quantum states is a central problem in quantum many-body physics. The single-particle spectral function constitutes a fundamental quantity in this context, as it encodes both the dispersion relations of excitations and their associated spectral weights~\cite{Coleman_2015}. Experimentally, two of the most prominent probes of the single-particle spectral function are angle-resolved photoemission spectroscopy (ARPES) \cite{ARPES_Review}, providing momentum- and energy-resolved measurements of the electronic spectral function, and scanning tunneling microscopy (STM) \cite{STM_Basics}, probing the local electronic density-of-states (LDOS) with energy resolution. Both methods, which rely on the injection or ejection of charges, 
have been instrumental in advancing our understanding of strongly correlated phases of matter. An important example is the study of doped antiferromagnets, where ARPES and STM experiments have been pivotal in characterizing high-$T_c$ superconductors \cite{ARPES_highTc_review,STM_HighTc_Review, Hashimoto2014-uf}, revealing the structure of the superconducting gap \cite{ARPES_Pseudogap,STM_Disorder_Gap} and the pseudogap phase \cite{ARPES_FermiSurface_HighTc, Li2024}.

Quantum simulators such as optical lattices or Rydberg tweezer array~\cite{CSB_Chen_2023, TLL_Emperauger_2025,Whitlock_2017, Scholl_2022,Homeier2024, tJ_qiao2025}, are now  also powerful platforms for studying strongly correlated matter with microscopic control and resolution ~\cite{Bloch2012_review, Blatt2012_review, Browaeys2020_review}. 
The recent discovery of correlated phenomena in moiré materials~\cite{tang_simulation_2020} has stimulated extensive theoretical investigations of unconventional phases on triangular lattices~\cite{Batista2017,MoreraAttractionFrustration,BruunTriangular,MoreraHighTmagnetic,Davydova2023,LiangFuPseudogap,Vafek2023}. In parallel, experiments on moiré heterostructures have revealed kinetic (Nagaoka) ferromagnetism~\cite{ciorciaro_kinetic_2023} and kinetically induced metamagnetism~\cite{taoSpinPolaron}, pointing to the emergence of novel magnetic polarons. These developments have further motivated the exploration of quantum simulators in triangular lattices. Quantum gas microscopes in optical lattices have provided key insights into the formation of kinetic magnetism in the spin-balanced regime by imaging spin polarons~\cite{XuOLtriangular,Lebrat2024,Prichard2024}. However, the highly polarized regime has remained largely unexplored, primarily due to the inefficiency of cooling in the absence of scattering-assisted mechanisms. More recently, Rydberg tweezer arrays have opened a new route to studying correlated systems in highly polarized settings, demonstrating that itinerant holes can bind to magnons to form composite quasiparticles~\cite{mu_kinetic_bound_state}. Although these bound states have been directly imaged in real space, their quasiparticle properties—such as binding energy and lifetime—remain largely uncharacterized, owing to the lack of suitable spectroscopic probes.

Tremendous efforts have been devoted to developing spectroscopic techniques for quantum simulation platforms~\cite{Stenger_1999,a_goerges_raman,Ness_2020,prichard_2025,Jurcevic_2015,cheng_quench}. 
Progress has been achieved in emulating ARPES using ultracold atoms in optical lattices \cite{Waseem_ARPES}. 
However, despite the existence of theoretical proposals \cite{GiamarchiSTM}, STM techniques are lacking in these platforms due to their limited ability to inject charges with simultaneous spatial and energy resolution. This limitation has prevented access to key characterizations of synthetic quantum systems, such as the binding energy of magnetic polarons.

Here we present a method  enabling the  measurement of  spectral functions in 
a Rydberg quantum simulator. It relies on the  frequency resolved injection of particles with adjustable spatial amplitudes. When injected on a single site, we recover STM-like measurements of the local Green's function. When the amplitudes of the injection has a standing wave profile, the spectral functions at a fixed momentum is similar to ARPES. 
Single-site resolved projective measurements of the many-body wavefunction gives access to equal-time multi-point correlation functions pivotal for characterizing nonlocal string order~\cite{Hilker2017} or topological phases~\cite{Sompet2022,Semeghini2021}. Combining this ability with spectroscopy in these platforms not only enables well-controlled analogs of ARPES and STM, but also extends beyond the conventional solid-state paradigm: Correlating spectral information with spatially-resolved snapshots, one can access more generalized correlation functions and achieve an unambiguous identification of quasiparticles contributions to spectral features.

We use a Rydberg quantum simulator to implement a $tJ$~model with hard-core bosonic holes in the strongly interacting limit ($J \rightarrow 0$), described by the Hamiltonian~\cite{Homeier2024, tJ_qiao2025}
\begin{equation}
\label{eq:tJ_Hamiltonian}
\begin{aligned}
&\hat{H}_{t} =\hbar\sum_{i<j}\sum_{\sigma=\downarrow,\uparrow
}t_{\sigma}\frac{a^3}{{r}_{ij}^3}\hat{\mathcal{P}}_{G}\left(\hat{b}_{i,\sigma}^\dagger\hat{b}_{j,\sigma}+\text { h.c. }\right)\hat{\mathcal{P}}_{G}\ .
\end{aligned}
\end{equation}
Here $\hat{b}^{\dagger}_{i,\sigma}$ is the creation operator for a hard-core boson with spin $\sigma$ at site $i$, and $\hat{\mathcal{P}}_{G}$ is the Gutzwiller projector that enforces the mutual hard-core constraint. 
The hole tunnels between neighboring sites, separated by $a$, with amplitude $t_{\sigma}>0$ resulting in kinetic frustration in triangular arrays \cite{mu_kinetic_bound_state}; in the following we set $t\equiv t_\downarrow$ and $t_\uparrow \approx 0.9 t_\downarrow$. 

Our spectroscopic method consists of injecting charges -- defined by the operator $\hat{h}_{i,\sigma}^{\dagger} = \hat{b}_{i,\sigma}$ and hereafter referred to as \emph{holes} -- using a local, time-dependent perturbation (see next section for the details on the modulation scheme) described by the Hamiltonian
in the rotating wave approximation:

\begin{align}\label{Eq:Hlocaldrive}
    \hat{H}(\tau) =\hbar \tilde{\Omega}\sum_{j} \alpha_je^{-i\omega\tau}\hat{h}^\dagger_{j\downarrow} + \mathrm{h.c.} 
\end{align}
with tunable energy~$E=\hbar\omega$, strength $\tilde{\Omega}$ and dimensionless local amplitudes~$\alpha_j$.
The transition rate for hole injection from a state with no hole initially is given at leading order in $\tilde{\Omega}$ by (see SM): 
\begin{equation}
    \Gamma = {d N_h\over d\tau}=2\pi \tilde{\Omega}^2 A(\omega),
    \label{Eq:GammaGF}
\end{equation}
where $N_h$ is the total number of holes and $A(\omega)=-\frac{1}{\pi}\mathrm{Im}\,G_h^{R}(\omega)$ is the single-hole spectral function related to the retarded hole Green’s function
\begin{align}
    G_h^{R}(\tau-\tau')=-i\theta(\tau-\tau') 
    \bra{0}   \hat{h}_{\downarrow}(\tau)  \hat{h}_{\downarrow}^{\dagger}(\tau')  \ket{0}, \label{Eq:RetardedGF}
\end{align}
with $\hat{h}_\sigma = \sum_j \alpha_j\hat{h}_{j,\sigma}$ and $\ket{0}$ designates a spin-polarized state with no hole; $\theta(\tau)$ is the step function.

Our protocol thus enables versatile, spectrally resolved charge injection: by choosing $\alpha_j = \delta_{j,j_0}$, holes can be injected locally at a single site $j_0$, realizing an atomic analogue of STM and enabling measurements of the local density of states (LDOS).
Alternatively, by choosing $\alpha_j = \cos(\vec{k}\cdot\vec{r}_j)$, holes are injected with a well-defined momentum $\vec{k}$, emulating ARPES. Such standing-wave-type light-shift modulations have been recently implemented in \cite{endres_CFT_spectroscopy}.

In the following, we first introduce and benchmark our method to characterize the systematic shifts induced by the probing. 
Then,  we use our technique to measure the binding energy between a hole and a magnon in a kinetically frustrated $tJ$ system~\cite{MoreraAttractionFrustration, mu_kinetic_bound_state}. Finally, we probe the LDOS in different 1d and 2d geometries.

\section{Light-shift-assisted local hole injection}

\begin{figure*}
\mbox{}
\includegraphics[width=1\textwidth]{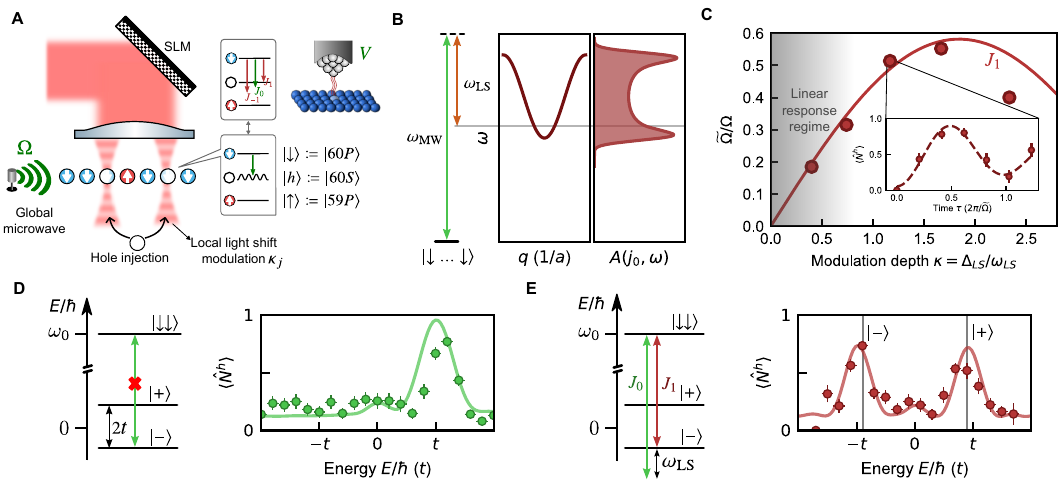}
\caption{\label{fig:sideband_spectroscopy_presentation} 
{\bf Measurement of spectral functions by local hole injection.}
\textbf{A,} Holes and spins are encoded in Rydberg states of each atoms, coupled by the dipolar interaction. The resulting many-body spectrum is probed with the combination of a global microwave and local modulations of the energy level encoding the hole.
\textbf{B,} The combined action of a global microwave and locally modulated light shifts induces the local injection of a hole at frequencies $\omega_{\rm MW}\pm \omega_{\rm LS}$. This local injection couples to all available states with arbitrary momentum $q$ at the corresponding energy (left panel). As a result, the protocol provides access to the local density of states (right panel).
\textbf{C,} Effective Rabi frequency of the oscillation (inset) driven by the first sideband on a single atom, as a function of the modulation depth $\kappa =\Delta_\mathrm{LS}/\omega_\mathrm{LS}$, compared to the expected Bessel function $J_1(\kappa)$ (solid line). 
\textbf{D,} Spectroscopy of a two-atom system with only a global microwave. The hole can only be injected at zero momentum and thus only the $\ket{+}$ state is probed.
\textbf{E,} Spectroscopy of the same system with one of the two site exposed to a modulated light shift. The resulting local sideband couples to both eigenstates. Vertical grey lines: expected positions of the $\ket{\pm}$ resonances. Solid lines: numerical simulations of an ideal system with an offset to account for detection and preparation errors.
Error bars correspond to one standard deviation.}
\end{figure*}

Our experimental setup relies on arrays of individual $^{87}$Rb atoms trapped in optical tweezers. We implement a hard-core bosonic $tJ$ model by encoding spin up, spin down, and hole states onto three Rydberg levels: $\ket{\downarrow} = \ket{60P_{3/2}, m_{J} = 1/2}$, $\ket{h} = \ket{60S_{1/2}, m_{J} = 1/2}$, $\ket{\uparrow} = \ket{59P_{3/2}, m_{J} = 1/2}$, see Fig.\,\ref{fig:sideband_spectroscopy_presentation}A. Resonant dipole interactions between atoms realize the bosonic $tJ$ Hamiltonian of Eq.\,\eqref{eq:tJ_Hamiltonian}. 

As the typical Rydberg transition frequencies lie in the $\mathrm{GHz}$ range, hence with a wavelength much larger than the extent of the tweezer array, a microwave field only injects particles (here holes) at zero momentum~\cite{Leseleuc2019}. Indeed, expressing in Fourier space a global microwave drive $\hat{H}_{\rm MW}(\tau)=\hbar\Omega e^{-i\omega_{\rm MW} \tau} \sum_j \hat{h}_{j,\downarrow}^{\dagger} + \textrm{h.c.}$, coherently coupling $\ket{\downarrow}$ to a hole state $\ket{h}$,  leads to $\hat{H}_{\rm MW}(\tau)=\hbar \sqrt{L}\Omega e^{-i\omega_{\rm MW} \tau} ( \hat{\tilde h}_{0,\downarrow}^{\dagger} + \textrm{h.c.})$, where $\hat{\tilde h}_{q,\downarrow}^\dagger=\frac{1}{\sqrt{L}}\sum_j e^{iqja}\hat{ h}_{j,\downarrow}^\dagger$ creates a hole with momentum $q$ in a system of size~$L$. Restricting ourselves to one hole, as we do in this work, this property is understood in real space by considering the matrix element between the initial polarized state 
$\ket{\downarrow...\downarrow}$ and the state with momentum $q$ containing one  hole, $\ket{\psi_q}=\frac{1}{\sqrt{L}}\sum_j e^{iqja}\ket{\downarrow...h_j...\downarrow}$. The global microwave has a vanishing matrix element to all finite momentum states: 
$\bra{\psi_q}\hat{H}_{\rm MW}\ket{\downarrow...\downarrow}\propto\sum_j e^{iqja}=\delta_{q,0}$. 
We therefore develop a two-photon technique to generate the local hole injection Hamiltonian of Eq.\,\eqref{Eq:Hlocaldrive}. 
It combines a global microwave drive $\hat{H}_{\rm MW}(\tau)$
with a spatially structured modulation of the light shift acting on the hole state of one or several atoms, $\hat{H}_{\rm LS}(\tau)=\hbar\Delta_{\rm LS}\cos(\omega_{\rm LS} \tau) \sum_j \alpha_j\hat{n}^h_j$, characterized by a global strength $\Delta_{\rm LS}$, a site-dependent function $\alpha_i$ and a modulation frequency $\omega_{\rm LS}$, with $\hat{n}_i^h=\sum_{\sigma}\hat{h}_{i,\sigma}^{\dagger}\hat{h}_{i,\sigma}$ being the density of holes. It is generated with a combination of blue- and red-detuned $1013\,\nm$ focused laser beams. The modulation is achieved with an acousto-optic modulator and the spatial profile is imprinted by a spatial light modulator (SLM) (see more details in the SM). The microwave is chosen far-detuned from all many-body states and atomic transitions.  
The modulation of  
the  energy of the site $j$ at a frequency $\omega_{\rm LS}$
results into a phase-modulation of the Rabi frequency at this site: in the weak modulation regime, it creates two sidebands at frequencies $\pm \omega_{\rm LS}$, with local amplitude $\alpha_j\widetilde\Omega =\alpha_j\Omega\,\Delta_{\rm LS}/2\omega_{\rm LS}$. 
When the microwave frequency is such that one of the sideband frequency $\omega_{\rm MW} \pm \omega_{\rm LS}$ matches a many-body excitation energy, a two-photon resonant transition takes place with a Rabi frequency $\widetilde\Omega \alpha_j$, as depicted in Fig.\,\ref{fig:sideband_spectroscopy_presentation}B.  The spatial control of these sideband amplitudes enables the coupling to one or all momentum states and realizes  the Hamiltonian \eqref{Eq:Hlocaldrive}.

Experimentally, to get a significant signal-to-noise ratio, we operate beyond the perturbative regime $\kappa=\Delta_\mathrm{LS}/ \omega_\mathrm{LS}\ll 1$. The local sideband amplitude is then given by $\alpha_j\Omega\,J_1(\kappa)$ ($J_n$ are the Bessel functions), 
as illustrated for a single atom ($\alpha_j=1$) in Fig.\,\ref{fig:sideband_spectroscopy_presentation}C. The tunneling amplitudes of the probed system are renormalized as $\widetilde{t}=J_0(\kappa)t$, with $J_0(\kappa) \rightarrow 1$ in the linear response regime (see more details in the SM).

\section{Two-atom demonstration}

We first demonstrate our protocol on two atoms encoding  one hole and 
the state $\ket{\downarrow}$. 
In this case, the eigenstates of the hole are the symmetric and antisymmetric combinations $\ket{\pm}=\left( \ket{h\downarrow}\pm\ket{\downarrow h}\right) / \sqrt{2}$ with energies $\pm \hbar t$ (see Fig.\,\ref{fig:sideband_spectroscopy_presentation}D,E). 
We first prepare the two atoms in the polarized state $\ket{\downarrow \downarrow}$, then measure the total number of holes $\hat{N}^h=\sum_j \hat{n}_j^h$ after the charge injection, i.e. the application of the microwave,
with and without the light shift modulation during $2.5$\,µs. 

Applying the microwave only and scanning $\delta_{\rm MW}$, 
we observe in Fig.\,\ref{fig:sideband_spectroscopy_presentation}D a peak at $E/\hbar = t$ corresponding to the coupling to 
symmetric state $\ket{+}$ with the matrix element
$\langle +|\Omega (\hat{h}_{1,\downarrow}^{\dagger}+
\hat{h}_{2,\downarrow}^{\dagger}+ \textrm{h.c.})\ket{\downarrow\downarrow}
=\sqrt{2}\,\Omega$. 
On the contrary the signal at 
$E/\hbar =-t$ corresponding to the state  $\ket{-}$ is absent, as
expected: the matrix element
$\langle -|\Omega (\hat{h}_{1,\downarrow}^{\dagger}+
\hat{h}_{2,\downarrow}^{\dagger}+ \textrm{h.c.})\ket{\downarrow\downarrow}=0$,
and the microwave alone cannot drive the zero momentum state
$\ket{\downarrow\downarrow}$ to the state $\ket{-}$ with momentum $\pi$. 

In order to probe this finite-momentum state, we apply our protocol 
and modulate a light shift applied on atom 1. The results of the 
microwave frequency scan is shown in Fig.\ref{fig:sideband_spectroscopy_presentation}E. We now observe a signal 
at frequency $\omega_{\rm MW}-\omega_{\rm LS}=-t$, indicating the 
coupling to the state $\ket{-}$. 

To quantitatively understand the data in Fig.\,\ref{fig:sideband_spectroscopy_presentation}E, we employ an expansion of the continuously driven two-atom Hamiltonian in powers of $\Omega/\LSmod$ and $t/\LSmod$ detailed in the SM (see also 
~\cite{James_2007, Goldman_2014, Goldman_2015}). This yields
\begin{equation}
\begin{aligned}
\label{eq:effective_zero_order_Natoms}
\hat{\widetilde H}
=
 -\delta_{\mathrm{eff}}
 (\hat{n}_1^h+\hat{n}_2^h)
 +\frac{\widetilde{\Omega}}{2}
 \hat{h}_{1,\downarrow}^{\dagger}
+\widetilde{t}\,  
 \hat{h}_{1,\downarrow}^{\dagger}
\hat{h}_{2,\downarrow}+ \textrm{h.c.}
\end{aligned}
\end{equation}
with $\delta_\mathrm{eff}=\delta_\mathrm{MW}-\LSmod$,
$\widetilde{\Omega}=J_1(\kappa)\Omega$ (with $\kappa=\Delta_{\rm LS}/ \omega_{LS}$) and a renormalized hopping term $\widetilde{t}=J_0(\kappa)t$. 
Experimentally, we use $\LSamp /h = 4$\,MHz and $\LSmod/h = 6$\,MHz, and operate beyond the perturbative regime with 
$\kappa\approx 0.7$. The  
probing thus affects the systems in two ways: firstly, it modifies the tunneling between the two sites by $J_0(\kappa)<1$; secondly, all off-resonant sidebands generate a light shift on the $\ket{\downarrow \downarrow}\leftrightarrow \ket{\pm}$ transition, with the carrier $J_0$ having the main contribution. The resulting shifts can be calculated theoretically. Besides we also take the  weak spin-spin (van der Waals) interactions into account. The position of the energies accounting for the probe and interaction induced shifts  are shown as gray dashed lines in Fig.\,\ref{fig:sideband_spectroscopy_presentation}E. They quantitatively match the positions of the experimental resonances.
Finally, the small peak observed around $E=0$ in 
Fig.\,\ref{fig:sideband_spectroscopy_presentation}D,E corresponds to the injection of two holes.

\section{Spectroscopy of a frustrated $tJ$ model}

\begin{figure}
\includegraphics[width=0.5\textwidth]{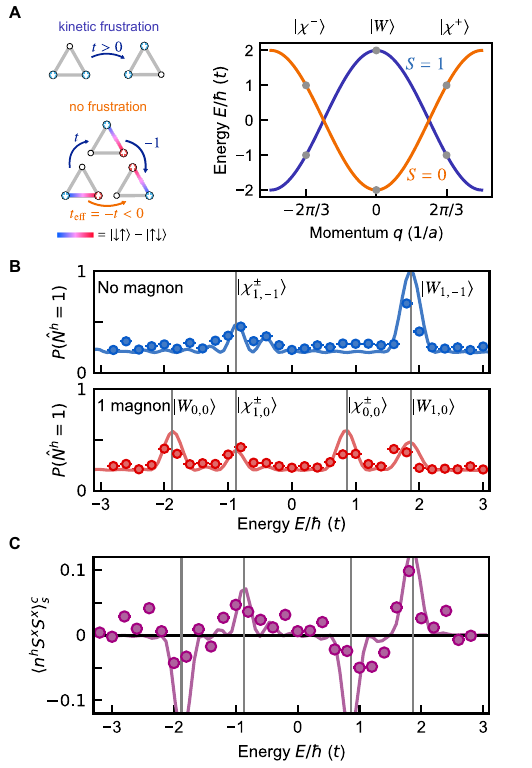}
\caption{\label{fig:3_atom_spectra} 
{\bf Sideband spectroscopy of a triangular plaquette.}
\textbf{A,} Mechanism of kinetically-induced
binding on a triangle. Left: A single hole with positive tunneling ($t>0$) experiences kinetic frustration. When a nearby magnon forms a singlet state with an adjacent spin, it effectively reverses the tunneling sign ($t^{\text{eff}} \approx -t < 0$). This relieves the frustration, lowering the kinetic energy. Right: Eigenstates and band structures of one hole hopping on 3 sites in a background forming a spin triplet (blue) or spin singlet (red).
\textbf{B,} Sideband spectroscopy of the triangular plaquette starting from $\ket{\downarrow \downarrow \downarrow}$ (top) or$\ket{\downarrow \downarrow h}$. Without magnon two peaks at $-t$ and $2t$ correspond to accessible triplet states. With one magnon additional peaks at $t$ and $-2t$ appears associated to singlet states; the latter is the origin of the hole-magnon bound state in triangular lattices.
\textbf{B,} Sideband spectroscopy with measurement along $x$ to reconstruct spin-spin correlations (see SM). Singlet (triplet) states showcase negative (positive) correlations. \textbf{B, C,} Solid lines: simulations of a perfect system with an offset to account for detection and preparation errors.
}
\end{figure}

We now exploit our protocol to study the kinetic frustration in a triangular $tJ$ model described by Eq.\,\eqref{eq:tJ_Hamiltonian}.
As discussed in the previous section, for a single bond, the hole lowers its kinetic energy by forming the antisymmetric wavefunction $\ket{-}$ between the two sites. On a triangular plaquette, however, this antisymmetric condition cannot be satisfied simultaneously across all three bonds, resulting in kinetic frustration~\cite{MoreraAttractionFrustration,OL_Prichard2024, mu_kinetic_bound_state}. The presence of a magnon $\ket{\uparrow}$ can relieve this frustration~\cite{Batista2017,MoreraAttractionFrustration}.

The eigenstates can be labeled by~$\ket{W_{S,m}}$ and~$\ket{\chi^{\pm}_{S,m}}$, where $S$ and $m$ corresponds to the total spin~$S=0,1$ and projection~$m$ of the spin background, and~$W$ ($\chi^{\pm}$) to the hole's momentum~$q=0$ ($q=\pm\frac{2\pi}{3}$). When the magnon forms an antisymmetric singlet state $S=0$, it effectively flips the sign of the hole's hopping amplitude ($t^{\text{eff}} = -t < 0$) and thus flips the  dispersion relation of the system $2t_\mathrm{eff}\cos(q)$, see Fig.\,\ref{fig:3_atom_spectra}A. With this negative effective hopping, the hole's kinetic energy is minimized by a symmetric wavefunction, which can be accommodated in the triangular plaquette without frustration. We experimentally illustrate this mechanism by performing the spectroscopy on three atoms arranged in a triangular plaquette, with and without a magnon, to highlight the apparition of new eigenstates with the spins forming a singlet state.

Similarly to the 2-atom case, we start from a polarized state, $\ket{\downarrow \downarrow \downarrow}$ or $\ket{\downarrow \downarrow \uparrow}$, apply the injection for a duration  $\tau=4$\,µs and measure the probability $P(\hat{N}_h=1)$ to get one hole. Spectra are obtained by varying $\delta_\mathrm{MW}$ so that the sideband $J_{-1}$ probes the different eigenstates. 
Starting from the state $\ket{\downarrow \downarrow \downarrow}$ we can reach the three eigenstates $\ket{W_{1,-1}}$ and $\ket{\chi^{\pm}_{1,-1}}$ with energies $2\hbar t$ and $-\hbar t$, respectively. The result of the experiment is shown in the top panel of Fig.\,\ref{fig:3_atom_spectra}B: we observe two peaks, one at $E=2\hbar t$ corresponding to the $q=0$ momentum state $\ket{W_{1,-1}}$, and a second one at $E=-2\hbar t$ for the two degenerate states $\ket{\chi^\pm_{1,-1}}$ of momentum $q=\pm 2\pi/3$ inaccessible by the microwave alone, due to the momentum conservation.

\begin{figure*}
\mbox{}
\includegraphics[width=1.\textwidth]{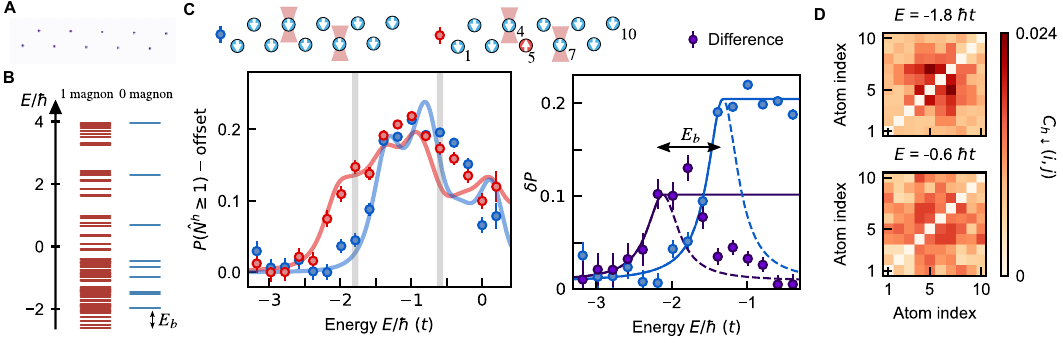}
\caption{\label{fig:binding_energy} {\bf Binding energy of a frustrated triangular ladder.}
\textbf{A,} Fluorescence image of a 10-site triangular ladder.
\textbf{B,} Eigenstates of the interaction Hamiltonian with (left) and without (right) a magnon. The presence of the magnon next to the hole relieves the frustration allowing the formation of  a hole-magnon bound state with binding energy $E_b$.
\textbf{C,} Left: Spectroscopy of a 10-atom triangular ladder for the two initial states. Initial configurations are illustrated above and show the two modulated sites. We subtract an offset corresponding to the baseline due to experimental errors, measured far-detuned from the spectrum. Solid lines correspond to a simple phenomenological model, see main text.  Right: Difference bewteen the two spectra on the left corresponding to the contribution of the bound states (purple) and the spectrum without magnon vertically shifted to remove the noise floor (blue). Both are fitted with a half-lorentzian (lines).
\textbf{D,} Measured hole-magnon correlations, obtained from postselecting on two missing majority spins~\cite{mu_kinetic_bound_state}, for two values of the detuning $\delta_\mathrm{MW}$ corresponding to $E=-1.8\hbar t$ (bound part of the spectrum) and  $E=-0.6\hbar t$ (continuum).}
\end{figure*}

As in the two-atom case, the probe induces small shifts of the resonances and their expected positions, accounting for them, are shown as gray vertical lines in Fig.\,\ref{fig:3_atom_spectra}B. The measurement does confirm that due to the kinetic frustration the triangular system can only reach a minimal energy of $-\hbar t$, higher than the $-2\hbar t$ that would be obtained in a frustration-free system.

We then apply the protocol starting from the product state $\ket{\downarrow \downarrow \uparrow}$, prepared by combining a MW pulse and a local light shift~\cite{CSB_Chen_2023,TLL_Emperauger_2025,tJ_qiao2025,mu_kinetic_bound_state}. 
The presence of the magnon in the initial state leads to six eigenstates: $\ket{W_{1,0}}, \ket{\chi^\pm_{1,0}}, \ket{W_{0,0}}, \ket{\chi^\pm_{0,0}}$. As seen in the lower panel of  Fig.\,\ref{fig:3_atom_spectra}B, the presence of the magnon results in two new peaks at energy $\hbar t$ and $-2\hbar t$ corresponding to the three singlet states. The peak at $-2\hbar t$ is a consequence of the $\hbar t$  gain in kinetic energy of the hole obtained by its pairing to a magnon. This coupling to the $\ket{W_{0,0}}$ state would not have been possible with the MW only.


To probe the formation of singlet and triplet states in the triangular plaquette, we measure the 3-body correlator $\langle n^h_iS^x_jS^x_k\rangle_{s}$ symmetrized on the permutation between different sites (see ~\cite{mu_kinetic_bound_state}). It corresponds to the spin-spin correlator conditioned on having injected a hole on the third site. Fig.\,\ref{fig:3_atom_spectra}C presents the result when starting from $\ket{\downarrow \downarrow \uparrow}$. We again observe four peaks associated to the six eigenstates. The two triplet peaks at $2\hbar t$ and $-\hbar t$ have positive values, as a triplet $\ket{\uparrow \uparrow}_x - \ket{\downarrow \downarrow}_x$ in $x$-basis, has positive correlation along $x$. In contrast, the singlet states feature  negative correlations in all directions. We observe negative correlations at $\hbar t$ and $-2\hbar t$, consistent with the formation of a singlet spin state.

We now generalize the method to a  ladder of equilateral triangles (edge length $a\approx 15$~µm) containing 10 atoms, as represented in Fig.\,\ref{fig:binding_energy}A. In this case, kinetic frustration leads to the binding of a hole and a magnon, as predicted in \cite{MoreraAttractionFrustration} and recently observed in our previous work \cite{mu_kinetic_bound_state}. Fig.\,\ref{fig:binding_energy}B presents the eigenspectrum of the ladder without and with one magnon. We observe the hole-magnon bound states at the bottom of the spectrum. 

Similarly to the 3-atom case, we consider two different initial states $\ket{\downarrow ... \downarrow}$ or $\ket{\downarrow... \downarrow\uparrow\downarrow...\downarrow}$ (see top of Fig.\,\ref{fig:binding_energy}C).
Here we choose to modulate a second site to enhance the signal
since the coupling arises from modulated atoms. 
We restrict ourselves  to the bottom, i.e. $E<0$, of the spectrum around the expected energies of the kinetically-induced bound states. 
Fig.\,\ref{fig:binding_energy}C shows the probabilities of having injected at least one hole: as the state  preparation and detection errors result in the detection of several holes, we compute $P(\hat{N}^h\ge 1)$ to mitigate the effects of these errors compared to $P(\hat{N}^h= 1)$. 

The experimental spectral showcases a strong signal between $-1.5\hbar t$ and $0$, with significant differences when performed with and without magnon:
without magnon the hole injection probability rapidly drops since there is no state to couple to, while in the presence of one magnon the bound states result in a stronger injection probability around $E=-2 \hbar t$. We interpret this higher probability as a signature of the coupling to the bound states. 
We compare the experimental spectra to the theoretical ones obtained by calculating a sum of phenomenological lorentzian profiles centered at the calculated energies $E_k$, with  width $\gamma=0.7\hbar t$  and weighted by the square of the matrix elements of 
the Hamiltonian \eqref{Eq:Hlocaldrive} coupling the initial state to the probed state $k$. This method leads to a good agreement with the data.  

To better contrast the cases with and without magnon, we compute the difference between the two spectra in the right panel of Fig.\,\ref{fig:binding_energy}C. 
We estimate the hole-magnon binding energy defined as the difference between the ground state energies of the system without and with magnon~\cite{mu_kinetic_bound_state, MoreraAttractionFrustration}: it  corresponds to the energy gained by the hole when pairing to a magnon forming a singlet. In the decaying part of the spectra on the left, these two lowest energy states are the only ones that contribute to the signal. We thus fit these low energy parts by a half-Lorentzian. We deduce the two ground states energy from the peak position of the fits and compute the difference $E_{\rm b}=(-0.8\pm 0.1)\hbar t$ in reasonable agreement with the
theoretical binding energy of a 10-site ladder, $E_b^\mathrm{theo}=-0.6 \hbar t$, directly obtained from the eigenspectrum shown in Fig. \ref{fig:binding_energy}B.

To confirm that we indeed probe the bound states, we analyze the symmetric hole-magnon correlations $C_{h\uparrow}(i,j)=\braket{n^h_in^\uparrow_j}+\braket{n^h_jn^\uparrow_i}$ ~\cite{mu_kinetic_bound_state} for  $E=-1.8\hbar t$, in the bound state part of the spectrum, and  $E= -0.6\hbar t$ in the continuum. The corresponding correlation maps 
at the end of the hole injection are represented in Fig.\,\ref{fig:binding_energy}D. The correlations are stronger close to the diagonal for  $E=-1.8\hbar t$, indicating that the hole and the magnon tend to stay close to each other and form a bound state, while they do not contain clear features in the continuum. Therefore, our spectroscopic protocol not only enables the extraction of the binding energy of the hole–magnon bound state, but also allows to extract spatial distribution of the composite after photoinjection.

\section{Atomic scanning tunneling spectrocopy}
\label{sec:STS}

\begin{figure}[h!]
\centering
\includegraphics[width=0.5\textwidth]{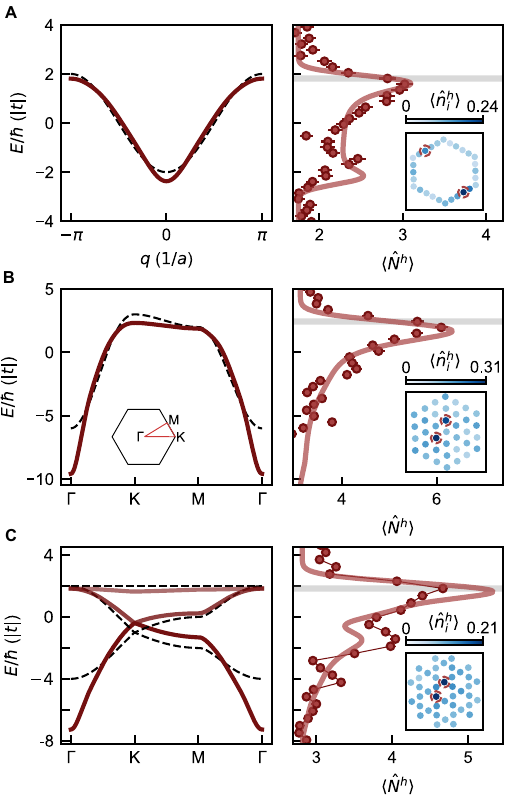}
\caption{\label{fig:vanHove} 
{\bf Atomic scanning tunneling spectroscopy.} 
{\bf A,} 1D ring, {\bf B,} triangular lattice, and {\bf C,} kagome lattice. 
{\bf Left panels:} Single-hole band structures for the different lattices calculated for the dipolar $1/r^3$ (solid lines)
and nearest-neighbor tunnelings (dashed lines). For the 2D geometries, a cut along the $\Gamma$-K-M points of the Brillouin zone is shown (see inset B, left). {\bf Right panels:} Red circles: Measured average number of holes $\langle \hat{N}^h \rangle$ as a function of the energy, which is proportional to the LDOS (error bars represent one standard error of the mean). We include a shift in the vertical axis to account for imperfect calibration. Insets show spatial maps of the atomic arrays where the color scale indicates the measured hole density $\langle \hat{n}^h_j \rangle$ at the energy shown by the gray line. The dashed circles identify the sites subjected to the local light-shift modulation, effectively acting as the "STM tip" for local hole injection. Solid lines represent theoretical LDOS from the band structure with a phenomenological Gaussian broadening to account for the probe's energy resolution. (A) shows an asymmetric density of states due to dipolar interactions; the triangular lattice (B) displays a singularity associated with saddle points at the M-points; and the kagome lattice (C) exhibits a strong peak at high energy, associated the the two peaks in the band structure due to finite energy resolution, and a peak at lower energy.  }
\end{figure}

As a final application of the method, we consider various lattice geometries and probe their single-particle bandstructure by exciting dilute holes into a polarized background, such that we can neglect hole-hole and spin interactions, up to a small global energy shift.
For large arrays, the energies 
are closely spaced and form a quasi-continuum. In this limit, our spectroscopic methods probes the LDOS of the holes $\rho(j,\omega)\propto \sum_n \langle \hat{n}^h_j  \rangle\delta(E_n-\hbar \omega)$. 
When the dispersion relation becomes sufficiently flat, the LDOS exhibits van Hove singularities~\cite{vanHove_1953}. While true singularities can be weakened in higher dimensions or in the presence of long-range tunnelings, the LDOS still features pronounced peaks. In Fig.\,\ref{fig:vanHove} left, we compare the nearest-neighbor dispersion (dashed) to the band structure with dipolar tunnelings (solid).

 
We investigate three different geometries, a 1d ring, a triangular lattice, and a kagome lattice, and use our spectroscopic method to probe the LDOS. To enhance the signal, we inject holes at two lattice sites, which--assuming a translationally invariant system with sufficiently large separation-- gives a signal that is twice as large.
In addition to increasing the system size, we also reduce the energy spacing by decreasing the hopping value and using a new mapping of the spin state $\ket{\downarrow} = \ket{60P_{3/2}, m_{J} = -1/2}$, leading to  $t/\hbar=-0.43$~MHz.
We start from an initial state where all the atoms are prepared in $\ket{\downarrow}$ and measure the density of holes~$\langle \hat{N}_h \rangle$ injected after $4\,\mu\mathrm{s}$ as a function of energy, which is proportional to the LDOS, see Eq.~\eqref{Eq:GammaGF}.

In the 1d ring, we observe a larger density of states close to the top of the band than at the bottom, see Fig.\,\ref{fig:vanHove}A. This asymmetry is a direct consequence of the $1/r^3$-scaling of the hopping, in contrast to the symmetric cosine dispersion for nearest-neighbour tunneling. In our case of a dipolar interaction, the hopping amplitudes constructively interfere for momentum $q=0$, making the dispersion relation steeper, while for $q=\pi$ they partially destructively interfere, leading to stronger singularities. Further, we analyze the single-site resolved density of particles, see insets in Fig.\,\ref{fig:vanHove} right, at the energy of the van Hove singularity. We observe a stronger injection of charges at the modulated sites. 
 
We then probe van Hove singularities in 2d systems. Unlike in 1d, where divergences occur at the band edges, in 2d, the density of states remains finite at the band extrema due to the volume elements entering in the integrals of the LDOS. Instead, the divergences arise from saddle points in the dispersion relation. For the triangular lattice, these saddle points are located at the M-points of the Brillouin zone~\cite{aoki2013physics}.
While the $1/r^3$ dipolar hopping modifies the width of the singularity,  these points are still degenerate in energy, resulting in a single singularity. This is what we observe in Fig.\,\ref{fig:vanHove}B. 

Finally, the Kagome lattice possesses three energy bands, leading to a richer density of states structure with three peaks. 
For nearest-neighbour tunneling, the two bands at lower energy are dispersive each hosting its own set of saddle points at the M-points, while the highest energy band is flat due to destructive interference. The dipolar tunnelings in our model give rise to a small bandwidth in the flat band, and further shift the saddle points to higher energies. In Fig.\,\ref{fig:vanHove}C, we find two large peaks: We associate the low-energy peak with one of the saddle points. The larger peak at high energies can be explained by the other saddle point and the high LDOS of the almost flat band, which both appear as a single peak due to the finite energy resolution in our experiment.

\section{Conclusion}

In conclusion, we have proposed and experimentally demonstrated a new spectroscopic technique that combines global microwave driving with local modulation to inject particles with both spatial and frequency resolution in a Rydberg quantum simulator. 
Our protocol effectively implements an atomic analogue of scanning tunneling spectroscopy, enabling direct measurements of the local density of states in a controlled environment across the entire relevant energy regime with exceptionally high frequency resolution. Through systematic benchmarks on small systems, we have developed a quantitative understanding of probe-induced energy shifts and the associated matrix elements. Furthermore, we have carried out the first spectroscopic determination of the binding energy of hole-magnon bound states.

Our work opens a new avenue for implementing generalized spectroscopic probes in Rydberg-encoded Hamiltonians. By extending the scheme to modulate all atoms with a standing-wave spatial profile, one could selectively couple to finite-momentum excitations set by the wavevector imprinted through the light-shift modulation. Such an approach would enable analogues of angle-resolved photoemission spectroscopy or inelastic neutron scattering, depending on whether holes or spin-flip excitations are injected, respectively. Adding a relative modulation phase between sites could also renormalize the tunneling by the first Bessel function similar to shaking in optical lattices. This would allow to implement synthetic gauge fields in Rydberg encoded-system~\cite{Goldman_2015}. Finally, we envision that our novel spectroscopic technique will offer new insights into the quantum simulation of exotic phases of matter using Rydberg tweezer arrays. Our atomic analogue of scanning tunneling spectroscopy can directly probe the emergence of pairing and its spatial structure in $tJ$-type Hamiltonians~\cite{ttprimeJ2025}. Moreover, this technique can be extended to measure the dynamical spin structure factor, thereby elucidating the nature of spin excitations. This approach offers a direct route to characterizing both the excitation gap and potential spin fractionalization in recently reported candidates for gapless spin liquids~\cite{bornet2026diracspinliquidcandidate}.\\

During completion of this work, we became aware of an
experiment implementing a related modulation spectroscopy method in a Rydberg quantum simulator \cite{endres_CFT_spectroscopy}.

\newpage
\clearpage

\subsection*{Supplemental Materials}

\textbf{Experimental setup.} Our Rydberg tweezer array experiment has been  described in previous works~\cite{TLL_Emperauger_2025, tJ_qiao2025, LS_Bornet_PRL}. For the experiment presented here, we have added a new addressing capability to apply a local modulation of the light shifts 
with zero mean. 

When a single beam addresses an atom, the resulting 
light shift is either positive or negative depending on the detuning of the laser beam with respect to the addressing transition. Here instead, we realize a locally addressable \emph{zero-mean} modulation by combining two $1013\,\mathrm{nm}$ beams detuned by $\delta_{\rm LS} = \pm 400$\,MHz with respect to the $6P_{3/2}-\ket{h}$ transition. The red- and blue-detuned beams of respective Rabi frequency $\Omega_O$ and $\Omega_M$ create a negative light shift $\Delta_O=-\Omega_O^2/4\delta_\mathrm{LS}$ and a postive one $\Delta_M=\Omega_M^2/4\delta_\mathrm{LS}$, see Supp. Fig.\,\ref{fig:SM_setup}A.

The two beams are combined by a polarizing beam splitter (PBS), then reflected from the same spatial light modulator (SLM) and share the same holographic phase pattern as shown in Fig.\,\ref{fig:SM_setup}B, which guarantees a common optical path and hence an excellent spatial overlap of the positive- and negative-shift patterns at the atomic plane. This overlap is essential: any relative misalignment would convert the intended purely time-dependent modulation into an unwanted static, site-dependent offset. We calibrate the contribution of each beam by addressing atoms with each laser separately and extracting the induced differential light shift from Ramsey measurements.


\begin{SM_fig*}
    \centering
    \includegraphics[width=\linewidth]{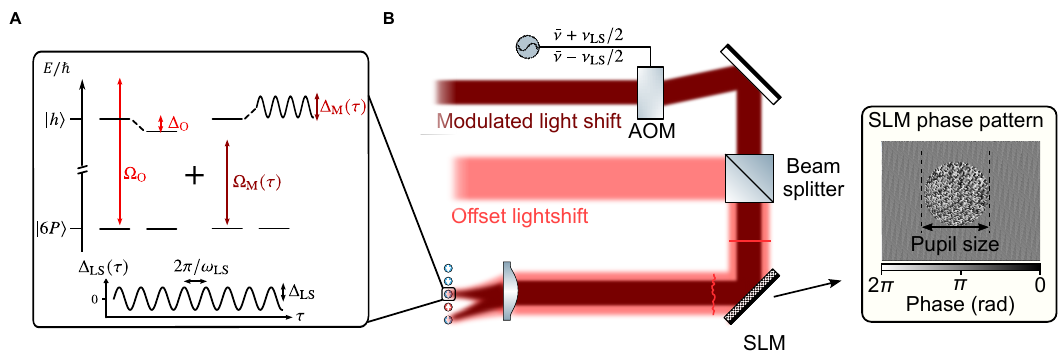}
    \caption{{\bf Experimental setup for generating the local  light-shift modulations.} \textbf{A,} Two far detuned beams (offset and modulated) are focused onto the atoms to generate an offset negative light shift and a modulated positive light shift. \textbf{B,} The two beams are combined at a beam splitter before reflecting off a SLM. The SLM imprints a holographic phase pattern with a defined pupil size to address specific atoms in the array with nearly uniform intensity. The AOM, driven by two frequency tones, generates the  time-periodic modulation of the light-shift potential.}
    \label{fig:SM_setup}
\end{SM_fig*}

To generate a periodic temporal modulation, the intensity of the red-detuned $1013\,\mathrm{nm}$ beam is modulated by an acousto-optic modulator (AOM) at frequency $\omega_{\mathrm{LS}}$ giving rise to a cosine positive light shift $\Delta_M(\tau)=\Delta_M(1+\cos(\omega_{\mathrm{LS}}\tau))$. By choosing the static negative shift to equal half of the peak positive shift $-\Delta_O=\Delta_M=\Delta_\mathrm{LS}$, the total light shift experienced by the addressed atom(s) takes the form
$\Delta_{\mathrm{LS}}(\tau)=\Delta_{\mathrm{LS}}\cos(\omega_{\mathrm{LS}}\tau)$,
i.e., a purely oscillatory modulation with vanishing time average. 
This minimizes unwanted global detuning drifts while preserving a well-controlled modulation depth. The set of parameters for the spectra shown in the main text are summarized in Table \ref{table:parameters}.\\

\begin{table}[ht]
    \centering
    \caption{Experimental parameters for the different spectra.}
    \label{table:parameters}
    \begin{tabular}{l c c c c}
    \toprule
    \textbf{Experiment} & $\Omega/h$\text{ (MHz)} & $\LSamp/h$\text{ (MHz)} & $\LSmod/h$\text{ (MHz)} & $\tau$\text{ (µs)} \\
    \midrule
    2-atom (a) & $0.2$ & $0$ 2 & $0$ 3 & $1.6$ \\
    2-atom (b) & $1$ & $4$ & $6$ & $2.5$ \\
    triangle & $1$ & $5$ & $8$ & $3$ \\
    10-atom ladder & $1$ & $5$ & $10$ & $4$ \\
    DOS spectra & $1$ & $5$ & $10$ & $4$ \\
    \bottomrule
    \end{tabular}
\end{table}

\textbf{The $tJ$ Model.} A comprehensive description of our implementation can be found in our previous work~\cite{Homeier2024, tJ_qiao2025}. Briefly, here we map the Rydberg states to a three-level system as follows: $\ket{\downarrow} = \ket{60P_{3/2}, m_{J} = 1/2}$, $\ket{h} = \ket{60S_{1/2}, m_{J} = 1/2}$, and $\ket{\uparrow} = \ket{59P_{3/2}, m_{J} = 1/2}$. The resonant dipole-dipole exchange and van der Waals (vdW) interactions between these states give rise to the $tJVW$ Hamiltonian:

\begin{widetext}
\begin{equation}
\begin{aligned}
\label{eq:Hamiltonian}
\hat{H}_{tJVW} = \hbar\sum_{i<j}\sum_{\sigma=\downarrow,\uparrow} & \left\{ t_{\sigma}\frac{a^3}{r_{ij}^3} \hat{\mathcal{P}}_{G} \left( \hat{b}_{i,\sigma}^\dagger\hat{b}_{j,\sigma} + \text{h.c.} \right) \hat{\mathcal{P}}_{G} \right. \\
& \left. + \frac{J^z}{r^6_{ij}} \hat{S}^z_i\hat{S}^z_j + \frac{V}{r^6_{ij}}\hat{n}^{h}_i \hat{n}^{h}_j + \frac{W}{r^6_{ij}} \left( \hat{S}^z_i \hat{n}^{h}_j + \hat{n}^{h}_i\hat{S}^z_j \right) + \frac{h^z_j}{r^6_{ij}}\hat{S}^z_j - \frac{\mu_j}{r^6_{ij}}\hat{n}^{h}_j \right\}
\end{aligned}
\end{equation}
\end{widetext}
where $\hat{S}_{j}^{z}$, $\hat{S}_{j}^{\pm} = \hat{S}_{j}^{x} \pm i \hat{S}_{j}^{y}$ are spin-1/2 operators for the states $\ket{\downarrow},\ket{\uparrow}$ at site $j$.
The hopping terms, $t_{\downarrow}$ and $t_{\uparrow}$ result from the  resonant dipole interactions, and scale as $1/r^3$. In contrast, the second-order dipole exchange~\cite{spin_exchange_gabriel} leads to an $XY$-type spin interaction $J_{\perp}$ that scales as $1/r^6$. The remaining term -- —the Ising-type spin interaction $J_{z}$, the spin-hole interaction $W$, the hole-hole interaction $V$, the on-site level shift $h_z$, and the boundary chemical potential $\mu$ -- originate from van der Waals interactions and all follow a $1/r^6$ scaling.

The quantization axis is given by a magnetic field~$B=46$~G pointing perpendicular to the atomic plane leading to spatially isotropic interactions. Here, we set the interatomic distance to $a = 14.7\,\si{\micro\meter}$ to suppress the $1/r^6$ terms. Consequently, the tunneling $t = t_\downarrow \approx 1.1 t_\uparrow$ becomes the dominant interaction, with all other contributions remaining below $\sim 5\%$, as summarized in Table~\ref{table:tJVW}.\\

\begin{table}[ht]
    \centering
    \caption{Interaction strengths of the various terms of the $tJVW$ model as implemented in the experiment.}
    \label{table:tJVW}
    \begin{tabular}{l S[table-format=-4.1] @{\hspace{3em}} l S[table-format=-2.1]}
    \toprule
    \textbf{Term} & \text{Strength (kHz)} & \textbf{Term}  & \text{Strength (kHz)} \\
    \midrule
    $t_{\uparrow}$    & -886.6  & $W$               & -2.9  \\
    $t_{\downarrow}$  & -1007.0 & $h_{z}$           & 1.6   \\
    $J_{\perp}$       & -1.9    & $\mu$             & -15.5 \\
    $J_{z}$           & -31.6   & $V$               & -10.8 \\
    \bottomrule
    \end{tabular}
\end{table}

\textbf{Perturbative spectroscopy.} We derive here Eqs.\,\eqref{Eq:Hlocaldrive} and \eqref{Eq:RetardedGF} of the main text, showing how 
they emerge from the combination of a global drive and local light-shift modulations. 
We assume first a weak light-shift modulation and treat it as a perturbation. We consider a of size $L$ governed by the $tJ$ model  subjected to the  external time-dependent perturbations $\hat H_{\text{ext}} = \hat H_{\text{MW}} + \hat H_{\rm LS}$ with (in this section we set~$\hbar\equiv 1$),
\begin{equation}
\begin{aligned}
&\hat H_{\text{MW}} = f_1(\tau)\hat V_1 e^{-i\omega_{\rm MW} \tau} + \text{h.c.},  \\&\hat H_{\rm LS} = 2f_2(\tau)\hat V_2 \cos(\omega_{\rm LS} \tau),
\end{aligned}
\end{equation}
where $\hat H_{\text{MW}}$ corresponds to a microwave pulse that uniformly creates a hole
$\hat V_1 =\sqrt{L} \Omega \hat h^\dagger_{q=0,\downarrow}$
at frequency $\omega_{\rm MW}$, and $\hat H_{\text{G}}$ corresponds to a modulation at frequency
$\omega_{\rm LS}$ of the local potentials of sites $j$ with spatial profile $\Delta^j_{\rm LS}$, giving
$\hat V_2 = \Delta_{\rm LS}\sum_{j} \alpha_j \hat{n}^h_j$.
Additionally, we introduce pulse envelopes $f_1$ and $f_2$ for each perturbation, respectively.

We want to calculate the rate of hole creation
$\Gamma = \frac{dN_h}{d\tau}$ 
with $N_h(\tau) = \langle \psi(\tau) | \hat N_h | \psi(\tau) \rangle$ the time-dependent number of holes and $|\psi(\tau)\rangle = e^{-i\hat H \tau}|\psi(\tau\to -\infty)\rangle $ the state after the perturbations are switched on. 

The time-evolution is initialized with no holes $\ket{0_0}$. We denote here $\ket{N_n}$ the eigenstate $n$ of $\hat{H}_{tJ}$ within the
manifold of states containing $N$ holes. By expanding the state in a perturbation series in terms of the interactions with the external potential $\hat H_{\text{ext}}$, we obtain, at second order,
\begin{align}
    |\psi(\tau)\rangle 
&= |0_0\rangle - i\int_{-\infty}^{\tau} d\tau_1\, \hat H_{\text{MW}}(\tau_1)|0_0\rangle \nonumber\\
&- \int_{-\infty}^{\tau} d\tau_2 \int_{-\infty}^{\tau_2} d\tau_1\, \hat H_{\rm LS}(\tau_2)\hat H_{\text{MW}}(\tau_1)|0_0\rangle .
\end{align}
The hole population $N_h^{(n)}(\tau)$ contains terms at second, third and fourth order in the external perturbations. Consider that the microwave is far-detuned from any many-body transition. Then, the action of the microwave alone does not lead to any significant change in the hole number and the first-order correction to the wavefunction can be neglected. Let us focus on the second-order correction to the wavefunction and its contribution to the transition rate at fourth order. We neglect the decay channel induced by the microwave field at third order. The transition rate is then given by:
\begin{widetext}
\begin{equation}
\label{Eq:Gamma4}
\begin{aligned}
\Gamma^{(4)}(\tau) &= 2\pi e^{2\eta \tau}
\sum_n
\int \frac{d\Omega}{2\pi}
|\tilde\mu_{n0}(\Omega)|^2
\frac{1}{\pi}
\left(
\frac{\eta}{(\omega_{\rm MW}+\omega_{\rm LS}+\Omega-E_{n0})^2+\eta^2}
+
\frac{\eta}{(\omega_{\rm MW}-\omega_{\rm LS}+\Omega-E_{n0})^2+\eta^2}
\right) \\
&\quad + 2\pi e^{2\eta \tau}
\sum_n
\int \frac{d\Omega}{2\pi}
\tilde\mu_{n0}(\Omega)
\frac{1}{\pi}
\left(
\frac{\eta\,\tilde\mu^{*}_{n0}(\Omega+2\omega_{\rm LS})}
{(\omega_{\rm MW}+\omega_{\rm LS}+\Omega-E_{n0})^2+\eta^2}
+
\frac{\eta\,\tilde\mu^{*}_{n0}(\Omega-2\omega_{\rm LS})}
{(\omega_{\rm MW}-\omega_{\rm LS}+\Omega-E_{n0})^2+\eta^2}
\right).
\end{aligned}
\end{equation}
\end{widetext}
where we introduce eigenenergy differences $E_{nm}=E_n-E_m$, effective matrix elements
$\tilde\mu_{n0}(\Omega)=
\sum_m \nu_{nm}\mu_{m0}F_{m0}(\Omega)$, with 
$$F_{m0}(\Omega)=
\int \frac{d\omega_{\rm MW}}{2\pi}
\frac{\tilde f_1(\omega_{\rm MW})\tilde f_2(\Omega-\omega_{\rm MW})}
{\omega_{\rm MW} + E_{m0}+i\Gamma_m/2}\ , $$ 
Here $\Gamma_m$ is the lifetime of state $m$, 
$\mu_{nm}=\bra{1_n} \hat{V}_2 \ket{0_m}$, and $\nu_{nm}=\bra{1_n} \hat{V}_2 \ket{1_m}$. The first line in Eq.\, \eqref{Eq:Gamma4} contains the co-rotating terms, and the second contains the counter-rotating ones. 

Let us now consider the case where the pulse envelopes vary 
very smooth with time $f_1 (\tau) = f_2 (\tau) \approx 1$. We then obtain a time-independent transition rate
\begin{widetext}
\begin{equation}
\Gamma^{(4)} =
2\pi \sum_n \left|\sum_m
\frac{\langle 1_n|\hat V_2|1_m\rangle
\langle 1_m|\hat V_1|0_0\rangle}
{\omega_{\rm MW}-E_{m0}+i\Gamma_m/2}\right|^2
\left[
\delta(\omega_{\rm MW}+\omega_{\rm LS}-E_{n0})
+
\delta(\omega_{\rm MW}-\omega_{\rm LS}-E_{n0})
\right].
\label{Eq:KH}
\end{equation}
\end{widetext}
Eq.~\eqref{Eq:KH} is reminiscent of the Kramers–Heisenberg (KH) formula, which describes second-order light–matter scattering processes and is widely used in solid-state spectroscopy, notably in resonant inelastic X-ray scattering (RIXS) \cite{RIXS_review}.

In the limit where the microwave detuning is much larger than all many-body excitation energies, $\omega_{\rm MW} \gg E_{m0}$, and the lifetime is long enough, $\omega_{\rm MW}\gg \Gamma_m$, the intermediate virtual states can be integrated out. The transition rate then simplifies to
\begin{widetext}
\begin{equation}
     \Gamma = -\frac{2\Delta^ 2_{\rm LS}\Omega^2}{\omega_{\rm MW}^2}{\rm Im} \left[G_h^{R}(\omega_{\rm MW} +\omega_{\rm LS}) + G_h^{R}(\omega_{\rm MW} -\omega_{\rm LS})\right],
     \label{Eq:GammaGF_SM}
\end{equation}
\end{widetext}
where ${\rm Im}[G_h^{R}(\omega)]$ is the single-hole spectral function related to the retarded hole Green’s function
 \begin{align}
     G_h^{R}(\tau-\tau')=-i\theta(\tau-\tau') 
     \bra{0_0}   \hat{h}_\sigma(\tau)  \hat{h}_\sigma^{\dagger}(\tau')  \ket{0_0}, \label{EqSI:RetardedGF}
 \end{align}
 with $\hat{h}_\sigma = \sum_j \alpha_j\hat{h}_{j,\sigma}$.
 Eq.\,\eqref{Eq:GammaGF_SM} shows that the two-photon process injects holes with a spatial profile determined by the light-shift modulation at frequencies $\omega_{\rm MW} \pm \omega_{\rm LS}$.

One way to validate the spectral-function measurement of our protocol consists in studying the scaling of the signal amplitude while simultaneously tracking the peak positions as a function of the drive amplitudes. One then extrapolates to the limit of small $\Omega \Delta_{\rm LS}/\omega_{\rm MW}$, where spectral-function measurements are reliable, with the signal amplitude expected to scale as $(\Omega \Delta_{\rm LS}/\omega_{\rm MW})^2$. However, in this weak-driving regime the signal can be difficult to resolve experimentally, motivating the consideration of higher-order effects.
Among them, an important one arises from the far-detuned microwave, which induces energy shifts analogous to  many-body AC Stark shifts. In the following, we present a non-perturbative treatment that incorporates these many-body energy shifts.\\

\textbf{Non-perturbative spectroscopy.} We now provide an alternative derivation in an Hamiltonian framework. We consider here the $tJ$ Hamiltonian in the strongly interacting limit:
\begin{equation}
\hat{H}_{t} = \hbar\sum_{i<j}\sum_{\sigma=\downarrow,\uparrow} t_{\sigma} \frac{a^3}{r_{ij}^3} \hat{\mathcal{P}}_{G} \left( \hat{b}_{i,\sigma}^\dagger \hat{b}_{j,\sigma} + \text{h.c.} \right) \hat{\mathcal{P}}_{G},
\end{equation}
As before, the system is subjected to a global microwave (MW) field with Rabi frequency $\Omega$ coupling  $\ket{\downarrow}$ to  $|h\rangle$ (transition frequency $\omega_0$), and a spatially patterned light shift (LS) modulation $\Delta_j \cos(\omega_{\text{LS}}\tau + \phi_j)$ acting on the hole density $\hat{n}_j^h = |h\rangle_j \langle h|_j$. The probe Hamiltonian  in the frame rotating at $\omega_{\rm MW}$ is thus:
\begin{align}
\label{eq:time_dependent_perturbation}
\hat{H}_\mathrm{mod}/\hbar=\LSamp\sum_{j}\alpha_{j}\cos(\omega_{\text{LS}}\tau+\phi_j)\hat{n}_{j}^{h}\nonumber\\
+\frac{\Omega}{2}\sum_{j} (\hat{d}^{\dagger}_{j}e^{-i\delta_{\text{MW}}\tau}+\text{h.c.}) 
\end{align}
where $\hat{d}_j^\dagger = |h\rangle_j \bra{ \downarrow}_j$ is the hole-creation operator, $\delta_{\text{MW}} = \omega_{\text{MW}} - \omega_0 $ is the microwave detuning, see Supp. Fig.\,\ref{fig:SM_bessel_benchmark}A.

\begin{SM_fig*}[hbt]
    \centering
    \includegraphics[width=\linewidth]{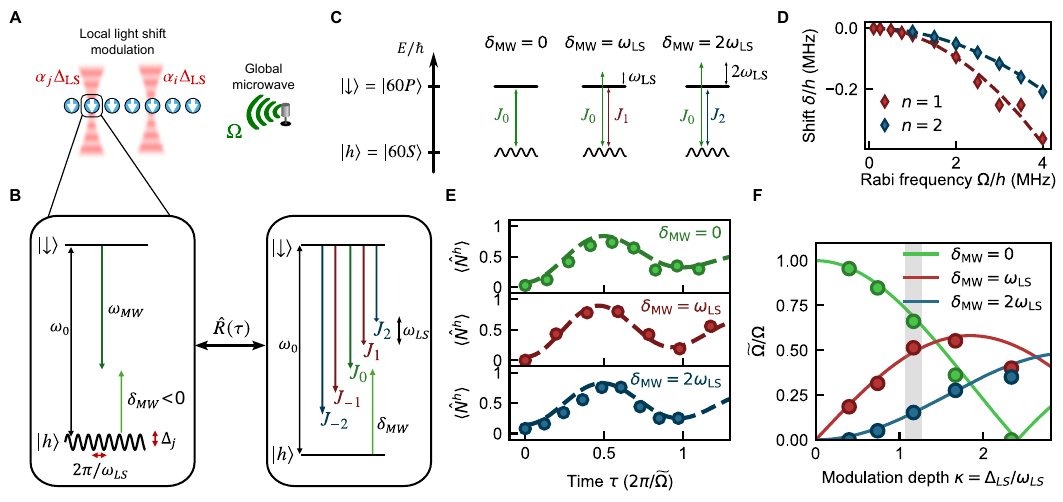}
    \caption{\textbf{Non-perturbative spectroscopy. A,} The system is subjected to a global microwave of Rabi frequency $\Omega$ and to some modulated light shift with local amplitudes $\alpha_i \LSamp$. 
    \textbf{B,} Modulating the energy of the site~$j$ at a frequency $\omega_{\rm LS}$ results into a phase-modulation of the Rabi frequency at this site: it creates sidebands at frequencies $n\omega_{\rm LS}$, with amplitude $\widetilde\Omega_j^n=\Omega\,J_n(\alpha_j\Delta_{\rm LS}/\omega_{\rm LS})$. 
    \textbf{C,} The microwave detuning $\delta_\mathrm{MW}$ can be chosen such that the $n$-th sideband is resonant with the transition. $\Omega\,J_j(\alpha_j\Delta_{\rm LS}/\omega_{\rm LS})$
    \textbf{D, } Light shifts obtained by simulating the time-dependent Hamiltonian given by Eq.\,\ref{eq:time_dependent_perturbation} for different values of the microwave Rabi frequency $\Omega$. The dotted lines correspond to the theoretical first order light shift deduced from Eq.\,\ref{eq:shift_expression}. Here $\omega_\mathrm{LS}/h = 20$MHz and $\Delta_\mathrm{LS}/h = 7$MHz.
    \textbf{E, }Examples of Rabi oscillations for a modulation depth of $\kappa= 1.17$ are shown for three different sidebands. The dashed line correspond to a fit by a damped cosine and errorbars corresponding to one standard deviation are always smaller than data symbols.
    \textbf{F, }Sidebands have relative amplitudes with the initial field given by Bessel functions leading to an effective Rabi frequency $\widetilde\Omega$ depending on the chosen sideband and the modulation depth $\kappa=\LSamp/\LSmod$. 
    \label{fig:SM_bessel_benchmark}}
\end{SM_fig*}


To make the effect of the local modulation more transparent, we follow the method introduced in~\cite{James_2007, Goldman_2015}. We first move to a modulated frame using the unitary transformation:
\begin{equation}
\hat{R}(\tau) = \exp \left[ i \sum_j \kappa_j \sin(\omega_{\text{LS}}\tau + \phi_j) \hat{n}_j^h \right],
\end{equation}
where $\kappa_j = \alpha_j\Delta_\mathrm{LS} / \omega_{\text{LS}}$ is the local modulation depth. In this modulated frame, the total Hamiltonian $\hat{H}=\hat{H}_t+\hat{H}_\mathrm{mod}$ transforms into $\hat{H}\rightarrow\hat{R}\hat{H}\hat{R}^{\dagger}-i\hat{R}\dot{\hat{R}}^{\dagger}$. As $[\hat{R}(\tau), \hat{n}_h]=0$, the hole density observable  keeps its original form in the new frame. In the new Hamiltonian, the energy of the hole  becomes  time-independent, 
but the hole-creation and hopping operators acquire time-dependent phases. A  decomposition  in Bessel functions $e^{i\kappa\sin\theta} = \sum_k J_k(\kappa)e^{ik\theta}$ leads to expression of the total Hamiltonian $\hat{H}$:
\begin{widetext}
\begin{equation}
\begin{aligned}
\label{eq:H1}
\hat{H}/\hbar = \sum_j \left[\sum_{k=-\infty}^{+\infty} \frac{\Omega J_{k}(\kappa_j)}{2} \left( e^{ik(\omega_{\text{LS}}\tau + \phi_j)-i\delta_{\text{MW}}\tau}\hat{d}_j^\dagger + \text{h.c.} \right) \right] + \sum_{i<j, \sigma} t_{ij, \sigma} \sum_{k=-\infty}^{+\infty}\hat{\mathcal{P}}_{G}\left[J_k(\kappa_{ji}) e^{ik\omega_{\text{LS}}\tau} \hat{b}_{i,\sigma}^\dagger \hat{b}_{j,\sigma} + \text{h.c.}\right]\hat{\mathcal{P}}_{G}
\end{aligned}
\end{equation}
\end{widetext}
where $\kappa_{ji} = \kappa_j - \kappa_i$ is the relative modulation depth. In this frame the frequency of the $\ket{\downarrow}\leftrightarrow\ket{h}$ is constant but the microwave is phase modulated, resulting in the apparition of local sidebands corresponding to the first term of this equation, as represented in Supp. Fig.\,\ref{fig:SM_bessel_benchmark}B.

We can select a particular $n$-th order of the Fourier sideband component to be resonant by setting $\delta_{\text{MW}}\sim n\omega_{\text{LS}}$, as illustrated in Supp. Fig.\,\ref{fig:SM_bessel_benchmark}C. It is then natural to move to the rotating frame of this $n$-th sideband with the unitary transformation $\exp{\left[i\delta_\mathrm{eff}\sum_j\sigma^z_j\right]}$ with $\delta_{\text{eff}}=\delta_{\text{MW}}-n\omega_{\text{LS}}$ the effective detuning. The total Hamiltonian becomes:
\begin{widetext}
\begin{equation}
\begin{aligned}
\label{eq:H1}
\hat{H}/\hbar = \sum_j \left[ -\delta_\mathrm{eff}\hat{n}^z_j+ \sum_{k=-\infty}^{+\infty} \frac{\Omega J_{k}(\kappa_j)}{2} \left( e^{i(k-n)(\omega_{\text{LS}}\tau + \phi_j)}\hat{d}_j^\dagger + \text{h.c.} \right) \right] + \sum_{i<j, \sigma} t_{ij, \sigma} \sum_{k=-\infty}^{+\infty}\hat{\mathcal{P}}_{G}\left[J_k(\kappa_{ji}) e^{ik\omega_{\text{LS}}\tau} \hat{b}_{i,\sigma}^\dagger \hat{b}_{j,\sigma} + \text{h.c.}\right]\hat{\mathcal{P}}_{G},
\end{aligned}
\end{equation}
\end{widetext}

which is time-periodic with a pulsation $\LSmod$. The effective dynamic is thus obtained by computing the time-average of this Hamiltonian which simply corresponds to the term $k=n$ in the sum of the previous equation. This leads to:
\begin{widetext}
\begin{equation}
\label{eq:tJ_model_effective}
\mathrlap{\hat{\widetilde{H}}}\phantom{H}^{}_{n}/\hbar = \sum_{j} \left[ -\delta_{\text{eff}}\,  \hat{n}_j^z + \frac{\widetilde{\Omega}^n_j}{2} \hat{d}_j^\dagger + \text{h.c.} \right] + \sum_{i<j, \sigma} \widetilde{t}_{ij, \sigma} \left( \hat{b}_{i,\sigma}^\dagger \hat{b}_{j,\sigma} + \text{h.c.} \right),
\end{equation}
\end{widetext}
with $\widetilde{\Omega}^n_j = \Omega J_n(\kappa_j)$ the effective local Rabi frequency and where the hopping is renormalized to $\widetilde{t}_{ij, \sigma} = t_{ij, \sigma} J_0(\kappa_{ji})$. We have thus neglected all the off-resonant sidebands ($k \neq n$ for the drive and $k \neq 0$ for the hopping). In the limit $\Omega/\LSmod\ll1$ and $t/\LSmod\ll1$, they are fast oscillating terms .


Beyond the resonant dynamics, the first-order correction in $\Omega/\omega_{\text{LS}}$ and $t/\LSmod$ is derived from the Fourier components of the total Hamiltonian of Eq.\,\eqref{eq:H1}, $\hat{H}=\sum_k\hat{H}^{(k)}e^{ik\LSmod}$. We obtain:

\begin{widetext}
\begin{equation}
\label{eq:many_body_effective_H_full}
    \mathrlap{\hat{\widetilde{H}}}\phantom{H}^{(1)}_{n}/\hbar = \sum_{k>0}\frac{[\hat{H}^{(k)},\hat{H}^{(-k)}]}{k\omega_{\text{LS}}\hbar^2}= \sum_j \delta_{\text{LS}, j} \hat{n}_j^h + \sum_{j \neq l} \sum_{k \neq 0} \frac{\widetilde{\Omega}^{n+k}_j \widetilde{t}_{jl}^k}{2k\omega_{\text{LS}}} \left( \hat{n}_j^h - \frac{1}{2} \right) \left( e^{in\phi_l} \hat{d}_l^\dagger + \text{h.c.} \right),
\end{equation}
\end{widetext}
where $\widetilde{\Omega}^{k}_j = \Omega J_k(\kappa_j)$ and $\widetilde{t}_{jl}^k = t_{jl} J_k(\kappa_{jl})$. The first term represents the AC Stark shift:
\begin{equation}
\label{eq:shift_expression}
    \delta_{\text{LS}, j} = \sum_{k>n} \frac{\widetilde{\Omega}^{(k)2}_j}{4(k-n)\omega_{\text{LS}}} - \sum_{k<n} \frac{\widetilde{\Omega}^{(k)2}_j}{4(n-k)\omega_{\text{LS}}},
\end{equation}
which shifts the energy of the hole state at site $j$ due to the presence of off-resonant microwave sidebands. Supp. Fig.\,\ref{fig:SM_bessel_benchmark}D compares the shifts given by this first order expansion with the ones obtained from the exact simulation of the dynamic with the time-dependent Hamiltonians of Eq.\,\eqref{eq:time_dependent_perturbation} with $\kappa=0.35$. Shift are very well captured by Eq.\,\eqref{eq:shift_expression} for $\Omega\ll\LSmod$. Together with the renormalization of the hopping $\widetilde t$, this correction explains the shifts of the peak positions observed in when the drive strength $\Omega$ or the modulation depth $\kappa$ is increased, moving beyond the simplest zeroth-order spectroscopic picture like in Fig.\,\ref{fig:sideband_spectroscopy_presentation}E and Fig.\,\ref{fig:3_atom_spectra}.

The second term in Eq.\,\eqref{eq:many_body_effective_H_full} represents a correlated second-order process where the creation of a hole at site $l$ is conditioned on the occupation state of site $j$. Physically, this term arises from virtual transitions involving one microwave-induced sideband off-resonant by $k\LSmod$ and one hopping-induced sideband also off-resonant by $k\LSmod$, making the second-order process resonant. Specifically, it couples the hole-free state $|0\rangle$ to a single-hole state $|\dots h_l \dots \rangle$ via an intermediate virtual state $|\dots h_j \dots \rangle$. The amplitude of this correction thus depends of the hole density of the final state at the modulated site. It results in a slightly different coupling to the different eigenstates. This correction is explicitly computed in the next section for the 2-atom system.
\\

\textbf{Single atom benchmark.} We benchmark the method on a single atom driving Rabi oscillations on the $\dspin-\ket{h}$ transition with different sidebands. The atom is prepared in  $\ket{\downarrow}$. We simultaneously modulate the $\ket{h}$-energy level and send a microwave detuned by $n \LSmod$ such that the sideband $J_n$ is resonant, see Supp. Fig.\,\ref{fig:SM_bessel_benchmark}C. The experiment is realized in a regime where $\LSmod \gg \Omega$ to ensure that only the resonant sideband contributes to the driving of the system. Rabi oscillations obtained for three different sidebands are shown in Supp.~Fig.\,\ref{fig:SM_bessel_benchmark}E. By fitting them, we extract their Rabi frequencies for different modulation depths $\kappa =\LSamp/\LSmod$ and find a good agreement with the theoretical Rabi frequency of the different sidebands $\widetilde{\Omega}=\Omega J_n(\kappa)$ (Supp. Fig.\,\ref{fig:SM_bessel_benchmark}F). We moreover find that these Rabi oscillations averaged over many realizations are characterized by a strong damping which we attribute to shot-to-shot  fluctuations of the light shifts and extension of the atom wavepacket. These fluctuations can modify the modulation amplitude $\LSamp$ but can also induce an offset if the positive and negative light shifts are not perfectly balanced. These two effects respectively result in a change of the effective Rabi frequency $\rabiEff$ and make the process slightly out of resonance, both causing a Gaussian damping  of the oscillations~\cite{spin_exchange_gabriel}. These fluctuations in the light shift originate mainly from the Gaussian spatial profile of the $1013\nm$ lasers as the atom experiences a different light shift depending on its postion. We reduce this source of decoherence by applying a pupil on the SLM phase pattern (see Fig.\,\ref{fig:SM_setup}B) in order to enlarge  the addressing beam waists, thus limiting light-shift fluctuations.\\ 

\textbf{Two-atom case.} We finally apply the non-perturbative spectrosocpy framework to a system of two interacting atoms, where only the first site is modulated ($\kappa_1 = \kappa$, $\phi_1 = 0,\kappa_2 = 0$). This results in a relative modulation depth $\kappa_{21} = \kappa$. We focus on the resonance condition for the first sideband ($n=1$), where the effective detuning $\delta_{\text{eff}} = \delta_{\text{MW}} - \omega_{\text{LS}}$ is small compared to the hopping strength.

In the modulated frame, the zeroth-order effective Hamiltonian $\mathrlap{\hat{\widetilde{H}}}\phantom{H}_{1}$ of Eq.\,\eqref{eq:tJ_model_effective} describes a system where the global microwave drive is converted into a local injection term at site 1:
\begin{equation}
\label{eq:2_atom_zero_order_H_new}
\mathrlap{\hat{\widetilde{H}}}\phantom{H}_{1}/\hbar = -\delta_{\text{eff}} (\hat{n}_1^h +\hat n_2^h)+ \frac{\widetilde{\Omega}}{2} \left(\hat{d}_1^\dagger + \text{h.c.} \right) + \widetilde{t}\, \left( \hat{h}_1^\dagger \hat{h}_2 + \text{h.c.} \right),
\end{equation}
where $\widetilde{\Omega} = \Omega J_1(\kappa)$ is the effective local Rabi frequency and $\widetilde{t} = t J_0(\kappa)$ is the renormalized hopping amplitude. The local nature of the drive ($\widetilde{\Omega}_2 = \Omega J_1(0) = 0$) explicitly breaks the exchange symmetry.

To evaluate the spectroscopic response, we write $\mathrlap{\hat{\widetilde{H}}}\phantom{H}^{(0)}_{1}$ in the single-hole manifold basis $\{\ket{h \downarrow}, \ket{\downarrow h}\}$. The eigenstates of the hopping term are  $\ket{\pm} = (\ket{h \downarrow} \pm \ket{\downarrow h})/\sqrt{2}$ with energies $E_{\pm} = \pm \hbar \widetilde{t}$. In this basis, the driving term $\hat{d}_1^\dagger$ transforms as:
\begin{equation}
\bra{\pm} \hat{d}_1^\dagger \ket{\downarrow \downarrow} = \frac{\widetilde\Omega}{2\sqrt{2}}.
\end{equation}
Consequently, both the symmetric $\ket{+}$ and antisymmetric $\ket{-}$ states are coupled to the initial state with an equal effective Rabi frequency $\Omega_{\text{eff}} = \widetilde{\Omega}/\sqrt{2}$, enabling the detection of the full energy spectrum.

The positions and strengths of these resonances are modified by the first-order corrections $\mathrlap{\hat{\widetilde{H}}}\phantom{H}^{(1)}_{1}$. For the two-atom system, the many-body AC Stark shift of Eq.\,\eqref{eq:shift_expression} acts only on the modulated site:
\begin{equation}
\label{eq:2atom_LS_final}
\delta_{\text{LS}, 1} = \sum_{k \neq 1} \frac{\Omega^2 J_k^2(\kappa)}{4(k-1)\omega_{\text{LS}}}.
\end{equation}
In the limit $\kappa \ll 1$, this shift is dominated by the $k=0$ (carrier), yielding $\delta_{\text{LS}, 1} \approx -\Omega^2/4\omega_{\text{LS}}$. Additionally, the correlated injection term of Eq.\,\eqref{eq:many_body_effective_H_full} introduces a correction to the coupling:
\begin{equation}
\label{eq:2_atom_cnot_new}
\mathrlap{\hat{\widetilde{H}}}\phantom{H}^{(1)}_1 = \sum_{i\ne j}\frac{\Omega \widetilde{t}J_1(\kappa_i) }{2k\omega_{\text{LS}}} \left( \hat{n}_{i}^h - \frac{1}{2} \right) \left( \hat{d}_{j}^\dagger + \text{h.c.} \right)\ ,
\end{equation}
where we have kept only the term $k=-1$ which dominates when $\kappa\ll1$. This term represents a two-photon process where a hole is virtually created at the modulated site and then hops to the neighbor. This process interferes with the direct sideband injection, leading to an asymmetry in the observed Rabi frequencies. Specifically, the effective coupling to the $\ket{-}$ state is enhanced while the coupling to $\ket{+}$ remains largely unaffected, as the virtual tunneling term flips the sign of the state $\ket{\downarrow h}$ thus $\bra{+}\mathrlap{\hat{\widetilde{H}}}\phantom{H}^{(1)}_{\text{corr}}\ket{\downarrow\downarrow}=0$ while $\bra{-}\mathrlap{\hat{\widetilde{H}}}\phantom{H}^{(1)}_{\text{corr}}\ket{\downarrow\downarrow}=\frac{\Omega J_1(\kappa) \widetilde{t}}{2k\omega_{\text{LS}}}$. The same results is obtained by a Taylor expansion of Eq.\,\eqref{Eq:KH} in $E_{-,0}=-\hbar t\ll \hbar \omega_\mathrm{MW}$.


\subsection*{Acknowledgements}
We thank Ana Maria Rey, Adam Kaufman and Martin Lebrat for fruitful discussions and feedback. The Institut d'Optique team acknowledges its CHADOQ apparatus for its support in this final project, ending 15 years of continuous operation and impressive scientific production.
This work is supported by the Agence Nationale de la Recherche (ANR-22-PETQ-0004 France 2030, project QuBitAF), and the European Research Council (Advanced grant No. 101018511-ATARAXIA), and the Horizon Europe programme HORIZON-CL4- 2022-QUANTUM-02-SGA (project 101113690 PASQuanS2.1)
R.M. acknowledges support by the ‘Fondation CFM pour la Recherche’ through a Jean-Pierre Aguilar PhD scholarship. L.H. acknowledges support by the Simons Collaboration on Ultra-Quantum Matter, which is a grant from the Simons Foundation (651440).
I.M. and E.D. acknowledge support by the SNSF (project 200021\_212899), the Swiss State Secretariat for Education, Research and Innovation (contract number UeM019-1) and the ARO (grant no. W911NF-20-1-0163).

\section*{Data Availability}

All data are available from the corresponding author on request.

\section*{Competing interests}
A.B. and T.L. are cofounders and shareholders of PASQAL. The remaining authors declare no competing interests.

\section*{Author contributions}
R.M., M.Q.,  I.M. and L.H. contributed equally to this work.
R.M. and M.Q. designed and carried out the experiments, with the help
of B.G., L.K., C.T., and D.B. I.M. and L.H. conducted the conceptual and theoretical analysis, and the numerical simulations. 
T.L., E.D. and A. B. supervised the work. All authors
contributed to the data analysis, progression of the project and on both the experimental and
theoretical sides. All authors contributed to the writing of the manuscript. 
Correspondence
and requests for materials should be addressed to Antoine Browaeys or Romain Martin.
\end{document}